\documentclass[aps,floatfix,twocolumn,superscriptaddress,showpacs]{revtex4}
\usepackage{amssymb}
\usepackage{amsbsy}
\usepackage{amsmath}
\usepackage{epsfig}
\usepackage[colorlinks,linkcolor=blue,citecolor=blue]{hyperref}

\makeatletter
\@ifundefined{textcolor}{}
{
 \definecolor{BLACK}{gray}{0}
 \definecolor{WHITE}{gray}{1}
 \definecolor{RED}{rgb}{1,0,0}
 \definecolor{GREEN}{rgb}{0,1,0}
 \definecolor{BLUE}{rgb}{0,0,1}
 \definecolor{CYAN}{cmyk}{1,0,0,0}
 \definecolor{MAGENTA}{cmyk}{0,1,0,0}
 \definecolor{YELLOW}{cmyk}{0,0,1,0}
 }
\makeatother

\begin{document}

\title{Quantum discord and quantum phase transition in spin-1/2 frustrated
Heisenberg chain}
\author{Chu-Hui Fan}
\affiliation{Department of Physics, Hangzhou Normal University, Hangzhou 310036, China}
\author{Heng-Na Xiong}
\affiliation{Department of Physics and Center for Quantum Information Science, National
Cheng Kung University, Tainan 70101, Taiwan}
\author{Yixiao Huang}
\affiliation{Zhejiang Institute of Modern Physics, Department of Physics, Zhejiang
University, Hangzhou 310027, China}
\author{Zhe Sun}
\email{sunzhe@hznu.edu.cn}
\affiliation{Department of Physics, Hangzhou Normal University, Hangzhou 310036, China}
\date{\today }

\begin{abstract}
By using the concept of the quantum discord (QD), we study the spin-1/2
antiferromagnetic Heisenberg chain with next-nearest-neighbor interaction.
Due to the $SU(2)$ symmetry and $Z_{2}$ symmetry in this system, we obtain
the analytical result of the QD and its geometric measure (GMQD), which is
determined by the two-site correlators. For the 4-site and 6-site cases, the
connection between GMQD (QD) and the eigenenergies was revealed. From the
analytical and numerical results, we find GMQD (QD) is an effective tool in
detecting both the first-order and the infinite-order
quantum-phase-transition points for the finite-size systems. Moreover, by
using the entanglement excitation energy and a universal frustration measure
we consider the frustration properties of the system and find a nonlinear
dependence of the GMQD on the frustration.
\end{abstract}

\maketitle

\section{Introduction}

Quantum correlation is one of the most popular research topics since it
plays a central role in quantum information and communication. Usually
people thought that the quantum computation devices should get their
computational power from entanglement---one of the most essential
nonclassical features in quantum mechanics. Recently, it was found that
there exist other nonclassical relations apart from quantum entanglement.
The concept of the quantum discord (QD) which is defined as the difference
between the quantum mutual information and the classical correlation was
introduced by Ollivier and Zurek~\cite{1} to quantify the non-classical
correlations. It has been observed that the QD provided a larger region of
quantum states with non-classical correlations, for example, even some
separable states have non-zero QD~\cite{2,3}. In fact, only zero discord is
a necessary condition for strictly classical correlations\thinspace ~\cite{4}%
, so that the states with non-zero QD are responsible for the efficiency of
a quantum computer~\cite{5}. Therefore, QD could be a new resource for
quantum computation, and even far cheaper and easier to maintain in the lab.

Quite recently, many people devoted into the study of quantum
discord\thinspace \cite{6,7,8,9,10,11,12,13,14,15,16,17,18,19}. Although
there are several advantages, a big problem for the application of discord
is that it is complicated to calculate QD analytically. Even for the
two-qubit systems, the analytic results of QD can only be obtained for a few
cases such as $X$-type states \cite{20}, and a general method still lacks.
Since that, Daki\'{c} \textit{et al. }\cite{20'} study the necessary and
sufficient condition for the existence of non-zero QD for bipartite states
and introduce a geometric measure of quantum discord (GMQD), which can be
evaluated for an arbitrary two-qubit state. The concept of the GMQD simplifies
the calculation of the QD.

In quantum many-body correlated system, quantum phase transition (QPT) is an
essential phenomenon and attracts widespread attention. As a quantum
critical phenomenon, QPT happens at absolute zero temperature, at which the
thermal fluctuations vanish so that no classical phase transition could
occur. Thus QPT is driven only by quantum fluctuation and caused by changing
the system's Hamiltonian, such as an external magnetic field or the coupling
constant. At the quantum phase transition point, the QPT behaves as the
configuration transition of the ground state (GS). Therefore, one can easily
expect that some quantum concepts closely related to the ground states can
be used to indicate QPTs. For example, the concepts of quantum
entanglement\thinspace \cite{22,23,24,25,26,27}, quantum fidelity\thinspace
\cite{28,29,30,31,32,32'} and quantum squeezing\thinspace \cite{34}, have
already been widely and successfully employed to study QPTs.

As a concept of quantum correlation, it is natural to consider the
relation between the QD and the QPTs. Recently, people become to make use of
QD to investigate QPTs. In Ref.\thinspace \cite{35}, it shows that QD
spotlights the QPT point for $XXZ$ Heisenberg chain even
at finite temperature. On the other hand, the quantum criticalities in the environment also play
an important role in the dynamics of the QD\thinspace \cite{36,37}. In some
typical systems with QPTs, it is found that both classical correlation and
quantum discord exhibit signatures of the QPTs\thinspace \cite{38}. All
these works imply that QD is an effective tool in detecting QPTs.

In this paper, in terms of the concept of QD, we will consider the QPTs of
the spin-1/2 Heisenberg chain with next-nearest-neighbor interaction which
is also called the $J_{1}$-$J_{2}$ model. This is an interesting quantum
many-body system for the existence of competition between nearest-neighbor
(NN) and next-nearest-neighbor (NNN) couplings. It well describes the
material structure in some quasi-one-dimensional compounds, such as CuGeO$%
_{3}$\thinspace \cite{J1-J2-1}. There are two important QPTs in this
model\thinspace \cite{J1-J2-2,J1-J2-3}, a first-order QPT and an
infinite-order QPT. The first-order QPT lies on the energy-level crossing of
the ground state (GS), while the infinite-order QPT is found to be closely
related to the energy-level crossing of the low-lying excited states (ESs)
and can be detected by the first-ES fidelity\thinspace \cite{32'}. However,
the properties of the QD in this system is lack of study, which motivates us
to consider and show that the QD of the GS and the ESs is effective to
indicate the QPTs in this system.

This paper is organized as follows: in Sec.\thinspace II, we introduce the
conception of the QD and the GMQD. In Sec.\thinspace III, we give a general
analytical result of the GMQD for the $X$-type states. In Sec.\thinspace IV,
we analytically calculate the eigenenergies of the $J_{1}$-$J_{2}$ model for
the 4-site and 6-site cases, then obtain the two-site GMQD and QD
analytically. In Sec.\thinspace V, we will numerically show that the GMQD is
a good candidate to indicate the QPTs. In Sec.\thinspace VI, we consider the
frustration by making use of the entanglement excitation energy (EXE) and a
universal measure $f$. Then we find the relation between frustration and the
GMQD. Finally, the conclusion will be given.

\section{Quantum discord and its geometric measure}

\subsection{Quantum discord}

Given a quantum state $\rho $ in a composite Hilbert space $\mathcal{H}=%
\mathcal{H}_{A}\otimes \mathcal{H}_{B}$, the total amount of correlation is
quantified by quantum mutual information~%
\begin{equation}
\mathcal{I}(\rho )=S(\rho _{A})+S(\rho _{B})-S(\rho ),
\end{equation}%
where $S(\rho )\equiv -\mathrm{Tr}\left[ \rho \log _{2}\rho \right] $ is the
von Neumann entropy and $\rho _{A(B)}=\mathrm{Tr}_{B(A)}\rho $ is the
reduced density matrix by tracing out system $B(A)$. If we take the system $%
A $ as the apparatus, the quantum discord is defined as follows~\cite{1,39}%
\begin{equation}
D(\rho )=\mathcal{I}(\rho )-\mathcal{C}_{A}(\rho ),
\label{eq:quantum_discord}
\end{equation}%
which is the difference of the total amount of correlation $\mathcal{I}(\rho
)$ and the classical correlation $\mathcal{C}_{A}(\rho )$. Here the
classical correlation is defined by
\begin{equation}
\mathcal{C}(\rho )=\max_{\{E_{k}\}}\mathcal{I}(\rho |\{E_{k}\}),
\label{eq:classical_correlation}
\end{equation}%
where $\mathcal{I}(\rho |\{E_{k}\})$ is a variant of quantum mutual
information based on a given measurement basis $\{E_{k}\}$ on system $A$ as
follows

\begin{equation}
\mathcal{I}(\rho |\{E_{k}\})=S(\rho _{B})-\sum_{k}p_{k}S(\rho _{B|k}).
\label{eq:variant_mutual_information}
\end{equation}%
$\rho _{B|k}=\mathrm{Tr}_{A}[(E_{k}\otimes \mathbb{I})\rho ]/p_{k}$ is the
postmeasurement state of $B$ after obtaining outcome $k$ on $A$ with the
probability $p_{k}=\mathrm{Tr}[(E_{k}\otimes \mathbb{I})\rho ]$. $\{E_{k}\}$
is a set of one-dimensional projectors on $\mathcal{H}_{A}$, and $\mathbb{I}$
is the $2\times 2$ identity operator.

\subsection{Geometric measure of the quantum discord}

In Ref.~\cite{20'}, Daki\'{c} \emph{et al.} proposed a geometric measure of
quantum discord defined by%
\begin{equation}
D_{g}(\rho ):=\min_{\chi \in \Omega _{0}}||\rho -\chi ||^{2},
\end{equation}%
where $\Omega _{0}$ denotes the set of zero-discord states and $||X||^{2}:=%
\mathrm{Tr}(X^{\dagger }X)$ is the Hilbert-Schmidt norm. For two-qubit
systems, a general state can be written in the Bloch representation \cite{20',40}%
:

\begin{equation}
\rho =\frac{1}{4}\mathbb{I} \otimes \mathbb{I}+\sum_{i}^{3}(x_{i}\sigma
_{i}\otimes \mathbb{I}+y_{i}\mathbb{I} \otimes \sigma
_{i})+\sum_{i,j=1}^{3}R_{ij}\sigma _{i}\otimes \sigma _{j}
\end{equation}%
with the real parameters $x_{i}$, $y_{i}$, and $R_{ij}$ , and the Pauli matrices $\sigma _{i=1,2,3}$.
Here we only consider the case that the measurement is taken
on the system $A$. Then an explicit expression of the GMQD is obtained
as\thinspace \cite{20'}:
\begin{equation}
D_{g}(\rho )=\frac{1}{4}\left( ||x||^{2}+||R||^{2}-k_{\mathrm{max}}\right) ,
\label{eq:GMQD_original}
\end{equation}%
where $x=(x_{1},x_{2},x_{3})^{T}$, $R$ is the matrix with elements $R_{ij}$,
and $k_{\mathrm{max}}$ is the largest eigenvalue of matrix $K=xx^{T}+RR^{T}$.

Here we introduce an alternative form which will be convenient when we
consider the evolution of the GMQD under decoherence. First, we introduce a
matrix $\mathcal{R}$ defined by\thinspace \cite{42}
\begin{equation}
\mathcal{R}=\left(
\begin{array}{cc}
1 & y^{T} \\
x & R%
\end{array}%
\right) ,  \label{R}
\end{equation}%
and another $3\times 4$ matric $\mathcal{R}^{\prime }$ obtained through
deleting the first row of $R$:
\begin{equation}
\mathcal{R}^{\prime }=(x,R).  \label{eq:R_prime}
\end{equation}%
Here $\mathcal{R}$ is just an expectation matrix with the elements $\mathcal{%
R}_{ij}=\mathrm{Tr}[\rho \sigma _{i}\otimes \sigma _{j}]$ for $i,j=0,1,2,3$,
and $\sigma _{0}=\mathbb{I}$ is defined. The expression (\ref{eq:R_prime})
leads to $K=\mathcal{R}^{\prime }(\mathcal{R}^{\prime })^{T}$. After
singular value decomposition, we have $\mathcal{R}^{\prime }=U\Lambda V^{T}$%
, where $U$ and $V$ are $3\times 3$ and $4\times 4$ orthogonal matrices, and
$\Lambda $ has only diagonal elements $\Lambda _{ij}=\mu _{i}\delta _{ij}$
with $\mu _{i}$ the so-called singular values of the matrix $\mathcal{R}%
^{\prime }$. Then the eigenvalues of the matrix $K$ can be expressed as $\mu
_{i}^{2}$. Considering $||x||^{2}+||R||^{2}=\mathrm{Tr}K$, we get an
alternative compact form of $D_{g}(\rho )$:
\begin{equation}
D_{g}(\rho )=\frac{1}{4}\left[ \left( \sum_{k}\mu _{k}^{2}\right)
-\max_{k}\mu _{k}^{2}\right] ,  \label{eq:GMQD}
\end{equation}%
where the summation and maximization are taken over all the non-zero
singular values $\mu _{k}$ of $\mathcal{R}^{\prime }$. This alternative form
will be convenient when we calculate the GMQD in the $J_{1}$-$J_{2}$ model.

\subsection{The Geometric measure of the quantum discord for $X$-state}

In the standard basis of operator $S_{z}=s_{1z}+s_{2z}$, the density matrix
of the so-called \textquotedblleft $X$-state\textquotedblright \ is shown as
follows:

\begin{equation}
\rho =\left(
\begin{array}{cccc}
a & 0 & 0 & g \\
0 & b & w & 0 \\
0 & w^{\ast } & c & 0 \\
g^{\ast } & 0 & 0 & d%
\end{array}%
\right) ,
\end{equation}%
where the parameters $a$, $b$, $c$ and $d$ are real numbers and satisfy $%
a+b+c+d=1$, and the positive condition requires $bc\geq \left\vert
w\right\vert ^{2}$ and $ad\geq \left\vert g\right\vert ^{2}$. Then we can
obtain parameter matrix $R$ as:

\begin{equation}
R=\left(
\begin{array}{ccc}
R_{11} & R_{12} & 0 \\
R_{21} & R_{22} & 0 \\
0 & 0 & R_{33}%
\end{array}%
\right) ,
\end{equation}%
where $R_{11}=2$Re$(g+w)$, $R_{12}=2$Im$(w-g)$, $R_{21}=-2$Im$(g+w)$, $%
R_{22}=2$Re$(w-g)$, and $R_{33}=a-b-c+d$. The parameter vectors $x%
=(0,0,x_{3})^{T}$ with $x_{3}=a+b-c-d$, and the vector $y=(0,0,y_{3})^{T}$
with $y_{3}=a-b+c-d$. By substituting the elements of $R$ and $x$ into
the matrix $\mathcal{R}^{\prime }=(x,R)$ (Eq.\thinspace (\ref%
{eq:R_prime})), we can calculate the GMQD by using Eq.\thinspace (\ref%
{eq:GMQD}) that
\begin{equation}
D_{g}(\rho )=\frac{1}{4}\left[ \mu _{1}^{2}+\mu _{2}^{2}+\mu _{3}^{2}-\max
\left( \mu _{1}^{2},\mu _{3}^{2}\right) \right] ,  \label{GMQD_X}
\end{equation}%
where $\mu _{1}^{2}=4\left( \left\vert g\right\vert +\left\vert w\right\vert
\right) ^{2}$, $\mu _{2}^{2}=4\left( \left\vert g\right\vert -\left\vert
w\right\vert \right) ^{2}$ and $\mu _{3}^{2}=2\left[ (a-c)^{2}+(b-d)^{2}%
\right] $. The expression above is the analytical result of the GMQD for the
$X$-states, which represent an important class of quantum states, e.g., the
$SU(2)$-symmetric states.

\section{GMQD for the Heisenberg model with NNN interactions}

The Hamiltonian of one-dimensional Heisenberg system with NNN interaction,
i.e., the $J_{1}$-$J_{2}$ model reads (let $\hbar =1$)
\begin{equation}
H=\sum \limits_{i}^{N}\left( J_{1}\mathbf{s}_{i}\cdot \mathbf{s}_{i+1}+J_{2}%
\mathbf{s}_{i}\cdot \mathbf{s}_{i+2}\right) ,  \label{Hnnn}
\end{equation}%
where the $\mathbf{s}_{i}$ denotes the spin-1/2 operator at the $i_{\text{th}%
}$ site. $N$ is the total number of sites and here we only consider the even
case. $J_{1}$ and $J_{2}$ are the nearest-neighbor (NN) and
next-nearest-neighbor (NNN) exchange couplings. As usual, we choose the
periodic boundary condition. Note that no exact analytical results are
available for this model (\ref{Hnnn}) except for the special case of $%
J_{2}/J_{1}=0$ and $J_{2}/J_{1}=1/2$. There are two important QPTs in this
model, a first-order QPT at $J_{c1}=0.5$ and an infinite-order QPT at $%
J_{c2}\simeq 0.241$ (where $J_{c}=J_{2}/J_{1}$).

At the critical point $J_{c1}=0.5$, the system reduces to the
Majumdar--Ghosh model\thinspace \cite{J1-J2-2}. Its GS is of spin zero, but
degenerate, which is a uniformly weighted superposition of the singlet
states between NN sites (for even and infinite N cases). This point is just
the GS energy-level crossing induced by the translation symmetry breaking,
thus it is a first-order QPT point.

At the other critical point $J_{c2}\simeq 0.241$, the system undergoes a
Berezinskii--Kosterlitz--Thouless (BKT) type QPT from spin fluid to
dimerized phase\thinspace \cite{J1-J2-3}. This phase transition is driven by
the competition between the NN and the NNN interactions. When $%
J_{2}/J_{1}<J_{c2}$, the NNN interaction does not change the character of
the simple antiferromagnetic case $J_{2}/J_{1}=0$, whose GS is described as
spin fluid massless spinon excitations. When $J_{2}/J_{1}>J_{c2}$, the
frustration term is relevant and the GS flows to the strong-coupling
dimerized phase. Furthermore, it is found that $J_{c2}$ is accurately the
energy-level crossing point of the first ESs for even-size and infinite-size
cases.
\begin{figure}[tbp]
\includegraphics[bb=-6 204 577 641,width=0.47\textwidth]{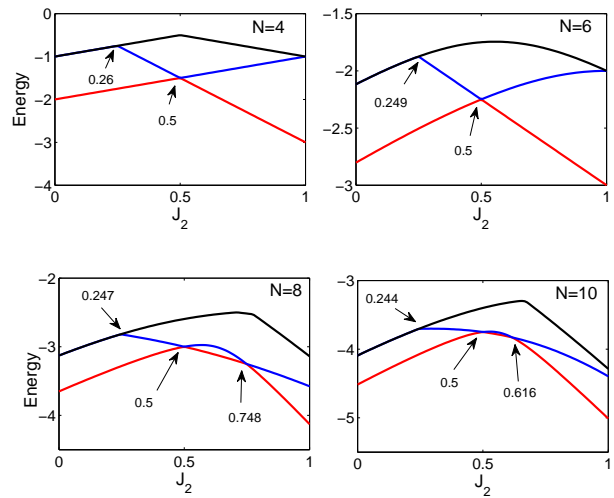}
\vspace{5pt}
\caption{Low-level energy spectrum of the spin-1/2 Heisenberg chain with NNN
interaction for different system sizes $N=4,6,8,10$. The parameters are in
units of $J_{1}$. }
\end{figure}

As shown in figure 1, we plot the low-energy levels of the GS and the first
and second ES versus different coupling $J_{2}$ (let $J_{1}=1$). Clearly,
for different sizes of $N=4,$ $6,$ $8,$ $10$, the GS energy-level crossing
occurs at the point $J_{c1}=0.5$, which exactly corresponds to the
first-order QPT point. Turning to the energy levels of the first ES, we find
that with the increasing system size, the crossing position moves from 0.25
to 0.244, which implies that in the thermodynamic limit, the energy-level
crossing point of the first ES will tend to the BKT QPT point $J_{c2}\simeq
0.241$. From the numerical results above, we emphasize that the QPTs are
closely related to the fundamental change of the energy structure.
Consequently, the corresponding eigenstates will present special quantum
properties at the QPT points.

Chen \textit{et al}~\cite{32'} studied the ground-state fidelity and
first-excited-state fidelity of this system. Xiong \textit{et al\thinspace }%
\cite{29} found the reduced fidelity can also be used to indicate the
quantum criticalities. In this paper, we will show that the two-site QD of
the GS and the first ES is also effective to detect the QPTs.

\bigskip Due to the $SU(2)$ symmetry and the $Z_{2}$ symmetry generated by
commutation $[\sigma _{x,y,z}^{\otimes ^{N}},H]=0$, the reduced density
matrix for the two-site subsystem $\rho _{ij}=$Tr$_{\overline{ij}}\left(
\rho _{\text{tot}}\right) $ with Tr$_{\overline{ij}}$ meaning tracing out
the subsystems except the $i$-th and $j$-th ones:
\begin{equation}
\rho _{ij}=\left(
\begin{array}{cccc}
a & 0 & 0 & 0 \\
0 & b & w & 0 \\
0 & w & b & 0 \\
0 & 0 & 0 & a%
\end{array}%
\right) ,  \label{reduced-density}
\end{equation}%
where all the elements are real numbers. Note that if the states are
degenerate, the reduced density matrices may not be in the form above. For
example, for the even-size $J_{1}$-$J_{2}$ Heisenberg system, the first ES
is threefold degenerate when $0<J_{2}/J_{1}<J_{c2}$ ($J_{c2}\simeq 0.241$),
and is nondegenerate when $J_{2}/J_{1}>J_{c2}$ except for the crossing
points. In order to overcome the above subtle problem induced by the
degeneracy, we mix the degenerate states with equal probability as that%
\begin{equation}
\rho _{n}=\frac{1}{G}\sum_{\upsilon =1}^{G}\left\vert \psi _{n\upsilon
}\right\rangle \left\langle \psi _{n\upsilon }\right\vert ,
\end{equation}%
where $G$ denotes the degeneracy of the energy $E_{n}$ and $\left\vert \psi
_{n\upsilon }\right\rangle $ the $\upsilon $-th degenerate eigenstate of the
system. This assumption is reasonable when we consider a general mixed state
in the thermal equilibrium $\rho (T)=\exp [-H/(k_{b}T)]/Z$, with $Z$ is the
partition funciton. If the eigenstate $\left\vert \psi _{n}\right\rangle $
of the Hamiltonian $H$ is degenerate, each of its degenerate states will
have an equal mixture weight in the mixed thermal state. It is easy to prove
that the equal-probability mixed state holds the $SU(2)$ and $Z_{2}$
symmetry, i.e., the reduced density is still in the form of Eq.\thinspace (%
\ref{reduced-density}).

By using the results of GMQD for $X$-state in Eq.\thinspace (\ref{GMQD_X}),
we obtain the GMQD for the reduced density matrix $\rho _{ij}$ as

\begin{equation}
D_{g}\left( \rho _{ij}\right) =\frac{1}{4}\left[ \mu _{1}^{2}+\mu
_{2}^{2}+\mu _{3}^{2}-\max \left( \mu _{1}^{2},\mu _{3}^{2}\right) \right] ,
\label{GMQD_J1J2}
\end{equation}%
where $\mu _{1}^{2}=4(a-b)^{2}$ and $\mu _{2}^{2}=\mu _{3}^{2}=4w^{2}$.

It is known that the elements of the reduced density matrix can be presented
by the expectation values of the Pauli matrices of the two-site subsystem,
i.e., $\left\langle \sigma _{i,\alpha }\sigma _{j,\beta }\right\rangle $
(with $\alpha ,\beta =x,y,z$ ).\ In addition, there is an exchange
invariance in the Hamiltonian, which leads to the fact that, any NN-site
correlators, i.e., $\left\langle \sigma _{i,\alpha }\sigma _{i+1,\beta
}\right\rangle $, equal to each other, so to the NNN-site correlators $%
\left\langle \sigma _{i,\alpha }\sigma _{i+2,\beta }\right\rangle $. Thus
when we consider the reduced density of the two-site subsystem, the elements
can be described by the correlators
\begin{eqnarray}
a &=&\frac{1}{4}\left( 1+\left\langle \sigma _{iz}\sigma _{jz}\right\rangle
\right) ,  \notag \\
b &=&\frac{1}{4}\left( 1-\left\langle \sigma _{iz}\sigma _{jz}\right\rangle
\right) ,  \notag \\
w &=&\frac{1}{4}\left( \left\langle \sigma _{ix}\sigma _{jx}\right\rangle
+\left\langle \sigma _{iy}\sigma _{jy}\right\rangle \right) ,  \label{abc}
\end{eqnarray}%
where $\sigma _{x,y,z}$ are Pauli matrices. In this system, there exists
another relation that $\left\langle \sigma _{i\alpha }\sigma _{j\alpha
}\right\rangle =\frac{1}{3}\left\langle \sigma _{i}\sigma _{j}\right\rangle $
with $\alpha =x,y,z$ and $\sigma _{i}\sigma _{j}=\sigma _{ix}\sigma
_{jx}+\sigma _{iy}\sigma _{jy}+\sigma _{iz}\sigma _{jz}$. Therefore, we can
prove that the three values $\mu _{1,2,3}$ in Eq.\thinspace (\ref{GMQD_J1J2}%
) equal to each other. Then the GMQD becomes
\begin{equation}
D_{g}\left( \rho _{ij}\right) =\frac{1}{2}\left\langle \sigma _{i\alpha
}\sigma _{j\alpha }\right\rangle ^{2}=\frac{1}{18}\left\langle \sigma
_{i}\sigma _{j}\right\rangle ^{2}.
\end{equation}%
This equation is a key result of this paper, and gives a general result for
the Heisenberg systems which contains the same symmetry properties of the $%
J_{1}$-$J_{2}$ model, such as $XXX$-type and dimerized Heisenberg chain.

Furthermore, for the $J_{1}$-$J_{2}$ model, by using of the Feynman-Hellman
theorem, one can find the correlators between the NN sites is
\begin{equation}
\left\langle \sigma _{i}\sigma _{i+1}\right\rangle =\frac{4}{J_{1}}\left[
e-J_{2}\frac{\partial e}{\partial J_{2}}\right] ,  \label{corrlator_energy}
\end{equation}%
where $e=\left\langle H\right\rangle /N$ denotes\ the expectation value of
the energy corresponding to each site. Similarly, for the NNN sites, we can
obtain the correlator $\left\langle \sigma _{i}\sigma _{i+2}\right\rangle
=4\partial e/\partial J_{2}$. Finally the NN-site GMQD will be connected to
the energy that
\begin{equation}
D_{g}(\rho _{i,i+1})=\frac{8}{9J_{1}^{2}}(e-J_{2}\frac{\partial e}{\partial
J_{2}})^{2},  \label{GMQD_energy}
\end{equation}%
and the NNN-site GMQD is
\begin{equation}
D_{g}(\rho _{i,i+2})=\frac{1}{18}\left\langle \sigma _{i}\sigma
_{j}\right\rangle ^{2}=\frac{8}{9}\frac{\partial e}{\partial J_{2}}.
\end{equation}%
The two equation above present the connection between the GMQD and the
energy. We can also give the QD of this model by using the results of \cite%
{16}%
\begin{eqnarray}
D(\rho _{ij}) &=&-2(a+b)\log _{2}(a+b)+2a\log _{2}a+2b\log _{2}b  \notag \\
&&+2w\log _{2}w-\frac{1}{2}[(1+4w^{2})\log _{2}(1+4w^{2})  \notag \\
&&+(1-4w^{2})\log _{2}(1-4w^{2})],
\end{eqnarray}%
by substituting the Eqs.\thinspace\ (\ref{abc}) and (\ref{corrlator_energy}%
), one can find the relation between the QD and the energy.

From the Eq.\thinspace (\ref{GMQD_energy}), we find that the GMQD is
dominated by the energy structure and its derivative. Obviously, the GMQD
will be sensitive to the singularity of the energy level. In other words,
the GMQD can effectively indicate the energy crossings. As is known, the
first-order QPTs usually lies on the energy-level crossing of the GS. Hence,
we can reasonably expect that the GMQD can indicate the first-order QPTs
effectively. In the $J_{1}$-$J_{2}$ model, there is another QPT point nearby
$J_{c2}\simeq 0.241$, which corresponds to the energy-level crossing of the
first ESs. Therefore, we can use the GMQD of the first ES to detect the
second QPT point $J_{c2}$.

In the following sections, we will analytically and numerically calculate
the small-size cases, which can also justify the fact that the GMQD can
indicate the QPTs in the $J_{1}$-$J_{2}$ model.

\section{Analytical results for the cases of $N=4$ and $N=6$}

In this section, we will analytically calculate the $J_{1}$-$J_{2}$ model
for 4-site and 6-site cases. The eigenenergy structure can be directly used
to obtain the GMQD. In the following, we set $J_{1}=1$ for simplicity.

Due to the periodic boundary condition, the Hamiltonian is translational
invariant, i.e., $\left[ H,T\right] =0$, where $T$ defines the cyclic
right-shift operator satisfying $T\left\vert
m_{1}m_{2}...m_{i-1}m_{i}\right\rangle =\left\vert
m_{i}m_{1}...m_{i-1}\right\rangle $.

\subsection{4-site case}

For $\left[ H,J_{z}\right] =0$, the 16-dimensional Hilbert space for a
4-site $J_{1}$-$J_{2}$ model can be divided into invariant subspaces spanned
by vectors with a fixed number of reversed spins\thinspace \cite{45}. Thus,
the dimension of the largest subspace is 6. With the help of translational
invariance, we can further reduce the Hamiltonian matrix to $2\times 2$
submatrices, and then the eigenvalue of Hamiltonian can be solved.

The none-reversed subspace contains only one vector $\left \vert
0000\right
\rangle $, which the eigenvalue is $E=1$. The subspace with one
reversed is spanned by 4 basis vectors, $\left \{ T^{n}\left \vert
1000\right \rangle ,n=0,1,2,3\right \} $. Considering the translational
invariance of the Hamiltonian, we choose another set of basis
vectors\thinspace \cite{46}:
\begin{equation}
\left \vert \Psi _{k}\right \rangle =\overset{3}{\underset{n=0}{\sum }}%
\omega _{k}^{n}T^{n}\left \vert 1000\right \rangle ,
\end{equation}%
where $\omega _{k}=e^{i2k\pi /4},k\in \{0,1,2,3\}$. One can check that $%
\left \vert \Psi _{k}\right \rangle $ is eigenstates of $T$ with eigenvalues
$\omega _{k}^{-1}$, and also are eigenstates for $H$ with eigenvalues $\frac{%
1}{2}(-1+\omega _{k}^{-1}+J_{2}\omega _{k}^{-2}+\omega _{k}^{-3}).$

When the reversed number is 2 or 3, we choose the basis%
\begin{equation}
\left\vert \Psi _{k}\right\rangle =\overset{3}{\underset{n=0}{\sum }}\omega
_{k}^{n}T^{n}\left\vert 1100\right\rangle ,
\end{equation}%
and
\begin{equation}
\left\vert \Phi _{k}\right\rangle =\overset{3}{\underset{n=0}{\sum }}\omega
_{k}^{n}T^{n}\left\vert 1010\right\rangle ,
\end{equation}%
for the 6-dimensional subspace. Under this space, one can rewrite the
Hamiltonian into a $2\times 2$ form and obtain all the eigenvalues. Then,%
\textit{\ }the GS energy is
\begin{equation}
E=\left\{
\begin{array}{cc}
-2+J_{2}\  & \text{ when }J_{2}\leq 0.5, \\
-3J_{2} & \ \text{when }J_{2}\geq 0.5.%
\end{array}%
\right.
\end{equation}%
$\bigskip $Substituting the energy levels above into Eq.\thinspace (\ref%
{GMQD_energy}), one can obtain the NN-site GMQD corresponding to the GS $%
\rho _{0}$:%
\begin{equation}
D_{g}(\rho _{0})=\left\{
\begin{array}{cc}
\frac{2}{9}\text{ } & \text{ when }J_{2}\leq 0.5, \\
0 & \text{ when }J_{2}\geq 0.5.%
\end{array}%
\right.
\end{equation}%
We also obtain the 1st ES energy:%
\begin{equation}
E=\left\{
\begin{array}{cc}
-1+J_{2}\  & \text{ when }J_{2}\leq 0.25, \\
-3J_{2}\  & \ \text{when }0.5\geq J_{2}\geq 0.25, \\
-2+J_{2}\  & \text{when}\ J_{2}\geq 0.5,%
\end{array}%
\right.
\end{equation}%
thus the NN-site GMQD corresponding to the first ES $\rho _{1\text{st}}$:

\begin{equation}
D_{g}(\rho _{1\text{st}})=\left\{
\begin{array}{cc}
\frac{1}{18} & \text{when }J_{2}\leq 0.25, \\
0 & \text{when }0.5>J_{2}\geq 0.25, \\
\frac{2}{9} & \text{when }J_{2}>0.5,%
\end{array}%
\right.
\end{equation}%
Obviously, the GMQD presents discontinuity at the point $J_{2}=0.5$, which
just corresponds to the first-order QPT of this system. While\ for the first
ES, the GMQD sudden changes at two points: $J_{2}=0.25$ and $J_{2}=0.5$.
Similarly, one can obtain the NNN-site GMQD, and we do not show the results
here.

\subsection{6-site case}

For the 6-site case, the energy structures are more complicated.
Fortunately, we analytically obtain all the eigenenergies. The GS energy is:%
\begin{equation}
E=\left\{
\begin{array}{cc}
-\frac{1}{2}\sqrt{9J_{2}^{2}-18J_{2}+13}-\allowbreak 1\  & \ \text{when }%
J_{2}\leq 0.5, \\
-\frac{3}{2}(1+J_{2}) & \ \text{when }J_{2}\geq 0.5.%
\end{array}%
\right.
\end{equation}%
Therefore, when $J_{2}\leq 0.5$ the NN-site GMQD for the GS is

\begin{equation}
D_{g}(\rho _{0})=\frac{1}{162}\left( \frac{18J_{2}-26}{\sqrt{%
9J_{2}^{2}-18J_{2}+13}}-\allowbreak 1\right) ^{2},
\end{equation}%
and when $J_{2}\geq 0.5$, we have a constant value of the GMQD: $D_{g}(\rho
_{0})=1/18$.

The first ES energy is:%
\begin{equation}
E=\left\{
\begin{array}{cc}
-\frac{1}{2}\sqrt{9J_{2}^{2}-10J_{2}+5}-1\  & \text{when}\ J_{2}\leq 0.25,
\\
-\frac{3}{2}(1+J_{2}) & \text{when }0.5\geq J_{2}\geq 0.25, \\
-\frac{1}{2}\sqrt{9J_{2}^{2}-18J_{2}+13}-\allowbreak 1 & \text{when }%
J_{2}\geq 0.5.%
\end{array}%
\right.
\end{equation}%
Thus when $J_{2}\leq 0.25,$ the NN-site GMQD is

\begin{equation}
D_{g}(\rho _{1\text{st}})=\frac{2}{81}\left( \frac{9J_{2}^{2}-19J_{2}+10}{2%
\sqrt{9J_{2}^{2}-10J_{2}+5}}+1\ \right) ^{2}.
\end{equation}%
When $0.5\geq J_{2}\geq 0.25$,\ \ we have a constant value of the GMQD that\
$D_{g}(\rho _{1\text{st}})=1/18$. When\ $J_{2}\geq 0.5$, the GMQD
becomes
\begin{figure}
\includegraphics[width=0.65\textwidth]{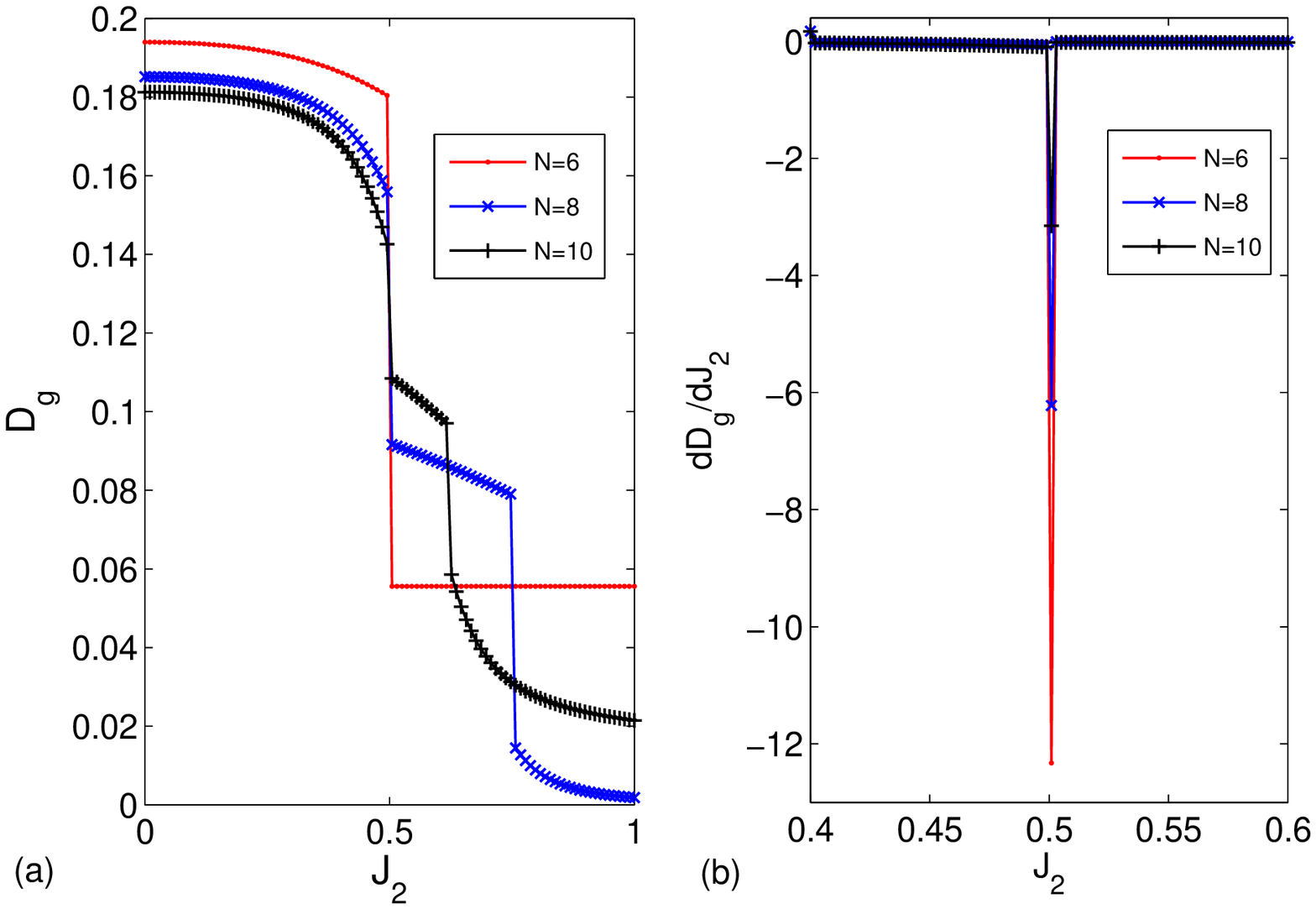}
\caption{(a) NN-site GMQD of the ground state versus coupling $J_2$ for
different system sizes $N=6,~8,~10$; (b) The derivative of the GMQD versus $%
J_2$. The parameters are in units of $J_1$.}
\end{figure}

\begin{figure}[tbp]
\includegraphics[width=0.65\textwidth]{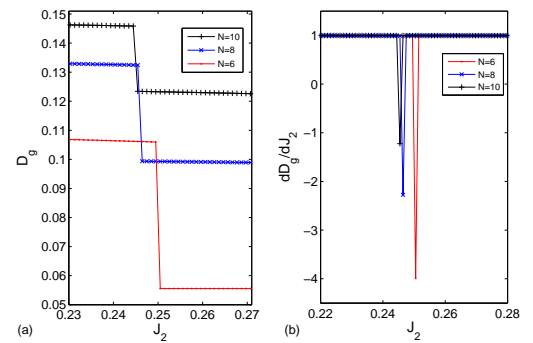}
\caption{(a) NN-site GMQD of the first excited states versus coupling $J_2$
for different system sizes $N=6,~8,~10$; (b)The derivative of the GMQD
versus $J_2$. The parameters are in units of $J_1$.}
\end{figure}
\begin{equation}
D_{g}(\rho _{1\text{st}})=\frac{2}{81}\left( \frac{9J_{2}^{2}-27J_{2}+22}{2%
\sqrt{9J_{2}^{2}-18J_{2}+13}}+\allowbreak 1\right) ^{2}.
\end{equation}%
Obviously, the GMQD of the GS changes at the point $J_{2}=0.5$, and the GMQD
of the ES changes at both the points $J_{2}=0.5$ and $J_{2}=0.25$.

From the analytical results of the GMQD, we see that the first-order QPT
point $J_{c1}=0.5$ can be indicated by the GMQD of the GS and ES. However,
for the small-size cases as 4 sites and 6 sites, the GMQD of the ES cannot
indicate the second QPT point $J_{c2}\simeq 0.241$. This is because the QPTs
occurs at the thermodynamic limit and the critical values is obtained in the
systems with infinite size. Although the small systems cannot accurately
reflect the quantum criticalities, we can simulate the critical properties
in the small systems\thinspace \cite{46,47}. In the following section, we
will numerically calculate the GMQD for some larger system with 8 sites and
10 sites. The second critical region can be approximately indicate by the
GMQD of the ES.

\section{ Numerical result}

In Fig.\thinspace 2a, we plot the NN-site GMQD of the GS versus coupling $%
J_{2}$ for different system sizes $N=6,8,10$. It is seen that the GMQD
presents discontinuity at the point $J_{c1}=0.5$. This is because that the
structure of the GS changes suddenly at the QPT point which can be reflected
by the quantum correlation, i.e., the GMQD. In Fig.\thinspace 2b, the QPT
point $J_{c1}=0.5$ is clearly detected by the derivative of the GMQD.

In order to study the BKT-type QPT, in Fig.\thinspace 3a we plot the NN-site
GMQD of the first ES versus coupling $J_{2}$ for different system sizes $%
N=6,8,10$. Obviously, there exists a sudden drop of the GMQD. Moreover, with
the increasing system size, the point of the drop tends to the QPT point $%
J_{c2}\simeq 0.241$. In Fig.\thinspace 3b, it is more obviously to find the
QPT point when we plot the derivative of the GMQD. The NNN-site GMQD of the
first ES is shown in Fig.\thinspace 4. Some different phenomena can be seen,
for the case $N=6,$ increasing coupling $J_{2}$ can enhance the NNN-site
GMQD, and the derivative of the NNN-site GMQD presents a positive peak. With
the increasing system size, the discontinuous position of the NNN-site GMQD
approaches the QPT point $J_{c2}\simeq 0.241.$

Note that as the system size increases, the discontinuous behavior around
the QPT point $J_{c1}=0.5$ and $J_{c2}\simeq 0.241$ becomes more and more
weak. One possible explanation is that for the QPTs associated with the
energy-level crossings, the continuity of the quantum correlation
characterized by the GMQD lies on the orthogonality of the eigenstates on
both sides of the point. However, in this paper we only consider the GMQD in
the reduced two-site subsystem where the orthogonality of the global system
is destroyed. Moreover, with the increasing system size, the orthogonality,
in other words, the information kept by the two-site subsystem occupies less
and less proportion of the global system. Therefore, the discontinuous
phenomena of the GMQD become weaker and weaker with the increasing system
size. Nonetheless, this GMQD approach for small finite-size systems is still
meaningful for the usual theoretical and experimental researches in the QPT
problems.

We also numerically calculate the QD (defined in Eq.\thinspace (\ref%
{eq:quantum_discord})) and find a similar discontinuous behavior with the
GMQD near the QPT points, thus we do not show the numerical results of the
QD in this paper. It is known that the GMQD can not absolutely characterize
the QD, e.g., under a local decoherence channel, a sudden change in the
decay rate of the GMQD does not always imply that of the quantum
discord\thinspace \cite{42}. However in this $J_{1}$-$J_{2}$ model, the GMQD
reflects the quantum criticalities as effectively as the QD. Furthermore,
the concise expression of the GMQD is more conducive to reveal the
connection between the quantum correlation and the eigenenergies. The
numerical calculation of the GMQD is more time-saving than that of the QD.
\begin{figure}[tbp]
\includegraphics[bb=-5 204 564 600,width=0.47\textwidth]{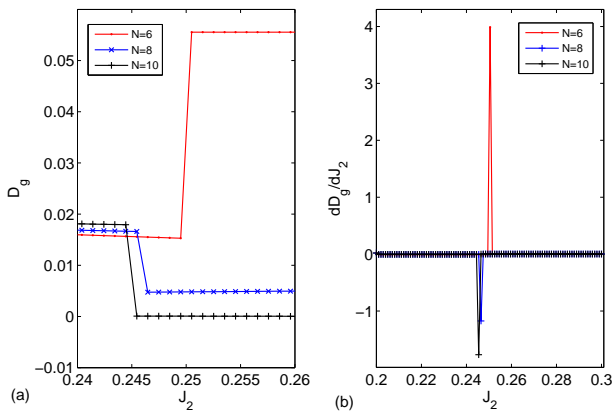}
\caption{(a) NNN-site GMQD of the first excited states versus coupling $J_2$
for different system sizes $N=6,~8,~10$; (b) The derivative of the GMQD
versus $J_2$. The parameters are in units of $J_1$.}
\end{figure}
\section{Frustration, GMQD and QPT}
As is konwn that the $J_{1}$-$J_{2}$ model characterizes a typical
frustrated spin system. Frustration arises from the simultaneous presence of
competing antiferromagnetic exchange interactions of different spatial
range\thinspace \cite{frustration-1}, and which causes the impossibility of
minimizing simultaneously the energy of competing interactions. Some methods
based on quantum information techniques were introduced to investigate
frustrations, e.g., the entanglement excitation energy (EXE) $\Delta E$%
\thinspace \cite{frustration-2}, and a universal measure of frustration $f$%
\thinspace \thinspace \cite{frustration-3}. In this section we will
analytically calculate the two quantities and find the relationship with
GMQD.

The entanglement excitation energy (EXE) is defined as
\begin{equation}
\Delta E=\min \left( \left\langle \Psi \right\vert U_{k}HU_{k}\left\vert
\Psi \right\rangle -\left\langle \Psi \right\vert H\left\vert \Psi
\right\rangle \right) ,
\end{equation}%
where $U_{k}=\otimes _{i\neq k}I_{i}\otimes O_{k}$ with $I_{i}$ denoting the
identity operator on all the spins but the one at site $k$, and $O_{k}$ is a
generic Hermitian, unitary, and traceless operator can be described as $%
O_{k}=$ $\sin \theta \cos \varphi \sigma _{k,x}+\sin \theta \sin \varphi
\sigma _{k,y}+\cos \theta \sigma _{k,z}$. If choosing the $\left\vert \Psi
\right\rangle $ as the GS the system $H$, we know that for any
translationally invariant and frustration-free Hamiltonian $H$ such that $%
[H,U_{k}]\neq 0$ $\forall U_{k}$, the vanishing of the EXE is a necessary
and sufficient condition for GS factorization\thinspace \cite{frustration-2}%
, and thus it is a proper measure of single-site entanglement. Indeed, the
presence of frustration tends to enhance correlations among the constituents
and thus to depress the possibility for the occurrence of separable
(uncorrelated) states. Hence entanglement and separability can be used to
qualify and quantify frustration\thinspace \cite{frustration-1}.

When calculating the EXE in the $J_{1}$-$J_{2}$ model, due to the $SU(2)$ and $Z
_{2}$ symmetry and also the translationally invariant, we have the
relations $\left\langle \sigma _{i,\alpha }\sigma _{j,\beta }\right\rangle
=0 $ for $\alpha \neq \beta $ and $\alpha ,\beta =x,y,z$, and $\left\langle
\sigma _{i,x}\sigma _{j,x}\right\rangle =\left\langle \sigma _{i,y}\sigma
_{j,y}\right\rangle =\left\langle \sigma _{i,z}\sigma _{j,z}\right\rangle $.
Finally we obtain that
\begin{equation}
\Delta E_{i}=-2J_{1}\left\langle \sigma _{i,z}\sigma _{i+1,z}\right\rangle
-2J_{2}\left\langle \sigma _{i,z}\sigma _{i+2,z}\right\rangle =-\frac{8}{3}%
e_{i},  \label{EXE}
\end{equation}%
where $e_{i}=\left\langle H\right\rangle /N$ denotes\ the expectation value
of the energy corresponding to each site. Especially in this model, no
matter what values of the parameters $\theta $ and $\varphi $ are chosen,
the EXE $\Delta E_{i}$ gives the same value. Obviously, $\Delta E_{i}$ can
be used to characterize the QPTs in the $J_{1}$-$J_{2}$ model due to the
direct relation with the energy level. The energy level crossing point also
corresponds to the sudden change of the EXE and thus indicate the QPTs. The
relation in Eq.\thinspace (\ref{EXE}) is still true in the excited states,
where $\Delta E_{i}$ may present negative values. Therefore, $\Delta E_{i}$
is effective to detect the second critical point $J_{2c}\approx 0.241$ when
taking into account the first excited state energy. Recalling the GMQD of
the NN spin pair in Eq.\thinspace (\ref{GMQD_energy}), then we have
\begin{equation}
D_{g}(\rho _{12})=\frac{1}{8J_{1}^{2}}(J_{2}\frac{\partial \Delta E_{i}}{%
\partial J_{2}}-\Delta E_{i})^{2},
\end{equation}%
and the GMQD of the NNN spin pair

\begin{equation}
D_{g}(\rho _{13})=\frac{1}{8}(\frac{\partial \Delta E_{i}}{\partial J_{2}}%
)^{2},
\end{equation}%
the two equations above connect the pairwise GMQD to the EXE and its
derivative. A vanishing EXE will induce a zero pairwise GMQD but the
converse may be not true. Finite EXE can also cause vanishing GMQD, because
GMQD characterizes the pairwise quantum correlations and not the single-site
ones. In other words, frustration can destroy the pairwise correlation and
hold the block correlation between single-site and the rest part.

\subsection{Universal measure of frustration and GMQD}

\begin{figure}[tbp]
\includegraphics[bb=11 215 584 652,width=0.45\textwidth]{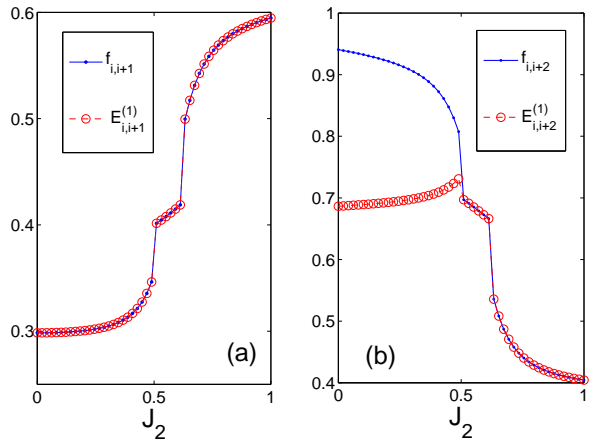}
\vspace{5pt}
\caption{(a) NN-site frustration measure $f_{i,i+1}$ and its lower bound $%
E^{(1)}_{i,i+1}$ versus coupling $J_2$ for the system size $N=10$; (b)
NNN-site frustration measure $f_{i,i+2}$ and its lower bound $%
E^{(1)}_{i,i+2} $ versus $J_2$ for the system size $N=10$. The coupling
strength $J_2$ is in units of $J_1$.}
\end{figure}

Recently a universal measure of frustration has been introduced as the
overlap between the global ground state of a frustrated model and the GS of
unfrustrated spin pairs\thinspace \cite{frustration-3}. For a many-body
Hamiltonian $H=\sum_{S}h_{S}$ with $h_{S}$ describing the local interaction.
The definition of the frustration corresponding to the subsystem $S$ is
\begin{equation}
f_{S}=1-\text{Tr}[\rho \Pi _{S}],
\end{equation}%
where $\rho $ denotes a pure state, and let $\Pi _{S}=\left\vert \Psi
\right\rangle _{s}\left\langle \Psi \right\vert $ be the projector onto the
subspace $S$ and $\left\vert \Psi \right\rangle _{s}$ is the GS of the local
interaction $h_{S}$ which is not frustrated. The quantity $f_{S}$ is a
well-defined measure of the interaction $h_{S}$. It turns out that the
geometric bipartite ground-state entanglement measured by
\begin{equation}
E_{S}^{(d)}=1-\sum_{i}^{d}\lambda _{i}^{\downarrow }\left( \rho _{s}\right)
\end{equation}%
is a universal lower bound to the frustration $f_{S}$, and $d$ is the rank
of the projector $\Pi _{S}$. For the degenerate GS $\left\vert \Psi
\right\rangle _{s}$, $d$ indicates the degeneracy. $\lambda _{i}^{\downarrow
}$ are the eigenvalues of $\rho _{s}=$Tr$_{R}\left( \rho \right) $ in
decreasing order.

In the $J_{1}$-$J_{2}$ model, the GS of $h_{S}=J_{1}\mathbf{s}_{i}\cdot
\mathbf{s}_{i+1}$ (or $J_{2}\mathbf{s}_{i}\cdot \mathbf{s}_{i+2}$) is the
singlet state $\left\vert \Psi _{-}\right\rangle =\left( \left\vert
10\right\rangle -\left\vert 01\right\rangle \right) /\sqrt{2}$, thus the
projector $\Pi _{S}=\left\vert \Psi _{-}\right\rangle \left\langle \Psi
_{-}\right\vert $ with $d=1$. Then we obtain the frustration measure is a
function of the correlator as:

\begin{equation}
f_{ij}=\frac{3}{4}+\frac{3}{4}\left\langle \sigma _{i\alpha }\sigma
_{j\alpha }\right\rangle ,\text{ }  \label{f_ij}
\end{equation}%
where $\alpha =x,y,z,$ and $j=i+1$ or $i+2$. \ \

From the frustration measure in Eq.$\,$(\ref{f_ij}), we exploit a nonlinear
relation with the GMQD that$\ $%
\begin{equation}
D_{g}\left( \rho _{ij}\right) =\frac{8}{9}\left( f_{ij}-\frac{3}{4}\right)
^{2},
\end{equation}%
from which, the GMQD achieves the maximum value $1/2$ when the frustration
vanishes $f_{ij}=0$, and has zero value when a finite frustration exists $%
f_{ij}=3/4$. This implies that the frustration in this model will depress
the pairwise correlation, however, it enhances the block correlation
between the two-site subsystem and the rest subsystem, which can be verified
by calculating the linear entropy
\begin{equation}
S_{l}\left( \rho _{ij}\right) =1-\text{Tr}\left( \rho _{ij}^{2}\right) =%
\frac{3}{4}-\frac{4}{3}\left( f_{ij}-\frac{3}{4}\right) ^{2},
\end{equation}%
obviously, when $f_{ij}=3/4$, the linear entropy has the maximum value which
signals the the maximal bipartite entanglement between the two-site
subsystem and the rest part of the system. To note that we now consider the
pure GS case, thus the linear entropy is an effective measure of the
bipartite\ entanglement and also the quantum correlation because we also
find a direct relation between the linear entropy and the GMQD that
\begin{equation}
S_{l}\left( \rho _{ij}\right) =\frac{3}{4}-\frac{3}{2}D_{g}\left( \rho
_{ij}\right) ,
\end{equation}%
from which one can clearly find that the maximal pairwise correlation $%
D_{g}\left( \rho _{ij}\right) =1/2$ will destroy the block correlation
between the two-site subsystem and the rest part, then $S_{l}\left( \rho
_{ij}\right) =0$.

The total measure of frustration is defined as $F=\frac{1}{M}%
\sum_{i,j}f_{ij} $, where $M$ is the number of bonds, and in our model $M=2N$
($N$ is the total number of the spin), then we obtain
\begin{eqnarray}
F &=&\frac{1}{2}\left( f_{i,i+1}+f_{i,i+2}\right)  \notag \\
&=&\frac{3}{4}+\frac{3}{8}\left( \left\langle \sigma _{i,\alpha }\sigma
_{i+1,\alpha }\right\rangle +\left\langle \sigma _{i,\alpha }\sigma
_{i+2,\alpha }\right\rangle \right)  \notag \\
&=&\frac{3}{4}+\frac{1}{2J_{1}}\left[ e_{i}+\left( J_{1}-J_{2}\right) \frac{%
\partial e_{i}}{\partial J_{2}}\right] ,
\end{eqnarray}%
in the above, we make use of $\left\langle \sigma _{i\alpha }\sigma
_{i+1,\alpha }\right\rangle =\frac{4}{3J_{1}}(e_{i}-J_{2}\frac{\partial e_{i}%
}{\partial J_{2}})$, and $\left\langle \sigma _{i\alpha }\sigma _{i+2,\alpha
}\right\rangle =\frac{4}{3}\frac{\partial e_{i}}{\partial J_{2}}$
corresponding to the NN sites and NNN sites respectively.

Now we calculate the lower bound
\begin{equation}
E_{ij}^{(1)}=\left\{
\begin{array}{c}
\frac{3}{4}+\frac{3}{4}\left\langle \sigma _{i\alpha }\sigma _{j\alpha
}\right\rangle \text{ when }\left\langle \sigma _{i\alpha }\sigma _{j\alpha
}\right\rangle \leq 0, \\
\frac{3}{4}-\frac{1}{4}\left\langle \sigma _{i\alpha }\sigma _{j\alpha
}\right\rangle \text{ when }\left\langle \sigma _{i\alpha }\sigma _{j\alpha
}\right\rangle >0.%
\end{array}%
\right.   \label{Ed_SR}
\end{equation}%
With the help of the energy levels shown in Fig.\thinspace 1, for\ the GS
energy level, obviously, one can find $\left\langle \sigma _{i\alpha }\sigma
_{i+2,\alpha }\right\rangle >0$ when $J_{2}<0.5$, and $\left\langle \sigma
_{i\alpha }\sigma _{i+2,\alpha }\right\rangle <0$ when $J_{2}\geq 0.5$. Thus
based on the Eq.\thinspace (\ref{Ed_SR}), we have
\begin{equation}
\left\{
\begin{array}{c}
f_{i,i+2}>E_{i,i+2}^{(1)}\text{ when }J_{2}<0.5, \\
\text{ }f_{i,i+2}=E_{i,i+2}^{(1)}\text{ when }J_{2}\geq 0.5.%
\end{array}%
\right.
\end{equation}%
It is known that the inequality $f_{ij}>E_{ij}^{d}$ implies the appearance
of the geometric frustration\thinspace \cite{frustration-3}. Hence, we find
that the geometric frustration appears when $J_{2}<0.5$ and vanishes when
when $J_{2}\geq 0.5$. It implies that the frustration measure and its lower
bound can reflect the change of the fundamental structure of the system,
such as the energy level crossing, therefore, they can be used to indicate
the first-order QPT point $J_{c1}=0.5$.

\bigskip Based on the analytical results of the GS energy level in the
4-site and 6-site case, we calculate the frustration measure. For 4-site
case (let $J_{1}=1$): the frustration of the NN sites
\begin{equation}
f_{i,i+1}=\left\{
\begin{array}{c}
\frac{1}{4}=E_{i,i+1}^{(1)}\text{ when }J_{2}<0.5, \\
\frac{3}{4}=E_{i,i+1}^{(1)}\text{ when }J_{2}\geq 0.5.%
\end{array}%
\right.
\end{equation}%
and the frustration of the NNN sites

\begin{equation}
f_{i,i+2}=\left\{
\begin{array}{c}
1>E_{i,i+2}^{(1)}=\frac{2}{3}\text{ when }J_{2}<0.5, \\
0=E_{i,i+2}^{(1)}\text{ when }J_{2}\geq 0.5.%
\end{array}%
\right.
\end{equation}%
For 6-site case, we have the frustration of the NN sites
\begin{equation}
f_{i,i+1}=\left\{
\begin{array}{c}
\frac{9J_{2}-13+7\Omega }{12\Omega }=E_{i,i+1}^{(1)}\text{ when }J_{2}<0.5,
\\
\allowbreak \frac{1}{2}=E_{i,i+1}^{(1)}\text{ when }J_{2}\geq 0.5,%
\end{array}%
\right.
\end{equation}%
where $\Omega =\sqrt{9J_{2}^{2}-18J_{2}+13}$, and the frustration of the NNN
sites
\begin{equation}
f_{i,i+2}=\left\{
\begin{array}{c}
\allowbreak \frac{3\left( \Omega -J_{2}+1\right) }{4\Omega }%
>E_{i,i+2}^{(1)}=\allowbreak \frac{3\Omega +J_{2}-1}{4\Omega }\text{ when }%
J_{2}<0.5, \\
\allowbreak \frac{1}{2}=E_{i,i+2}^{(1)}\text{ when }J_{2}\geq 0.5.%
\end{array}%
\right.
\end{equation}%
\

For larger size system such as $N=10$, we numerically calculate the
frustration measure and its lower bound in Fig.\thinspace 5. Both the
NN-site frustration measure $f_{i,i+1}$ in subfigure (a) and the NNN-site
frustration measure $f_{i,i+2}$ in subfigure (b) present a sudden change at
the QPT point $J_{2}=0.5$. Moreover, in the subfigure (a), we shows that the
NN-site frustration measure $f_{i,i+1}$ and its lower bound $E_{i,i+1}^{(1)}$
are consistent with each other, which implies that only quantum frustrations
exist between NN sites. Differently, in the subfigure (b), we find the
NNN-site frustration $f_{i,i+2}>E_{i,i+2}^{(1)}$ when $J_{2}<0.5$, which
signals the appearance of the geometric frustration.

From the above, we find the nonlinear dependence of the GMQD on the
frustration, which implies that the pairwise quantum correlation
characterized by the GMQD is greatly affected by the frustration, e.g., some
finite frustration as $f_{ij}=3/4$ can depress the\ GMQD to zero. On the
other hand, we believe that the frustration measure can be used to
characterize the first-order QPT in the $J_{1}$-$J_{2}$ model and also other
systems. In addition, the inequality $f_{S}\geq E_{S}^{(d)}$ holds as well
in any mixed states\thinspace \cite{frustration-3}, thus the method is also
effective to detect the first-order QPTs in the degenerate GSs. In our $%
J_{1} $-$J_{2}$ model, the quantity $f_{S} $ can also be used in the
first-excited states to detect the QPT point $J_{c2}\approx 0.241$. However,
in the excited states (ESs) case, the quantity $f_{S}$ can not be understood
as the frustration measure, instead, it only quantifies how much fails to
fully overlap with the subspace selected by the projector $\Pi _{S}$, and
the projector $\Pi _{S}$ may come from the GS or the ES of the local
interaction $h_{S}$.

\section{Conclusion}

In this paper, we considered the quantum discord (QD) in the Heisenberg spin
chain with next-nearest-neighbor (NNN) interaction. By using the geometric
measure of the quantum discord (GMQD), we studied the quantum correlation
properties of the ground states (GSs) and the first excited states (ESs).

We give a general analytical result of the GMQD for the $X$-type states. For
the Heisenberg system with the $SU(2)$ symmetry and $Z_{2}$ symmetry, we
give an exact relation between the GMQD and the two-site correlators.
Furthermore, the connection between GMQD and the eigenenergies was revealed.
For the 4-site and 6-site cases, the analytical results of the GMQD for the
GS and first ES are obtained, from which the first critical point $%
J_{c1}=0.5 $ can be exactly detected. We also numerically studied the
NN-site and NNN-site GMQD of the ESs. It is found that when the system size
increases from 6 sites to 10 sites, the discontinuous point of the GMQD
tends to the second QPT point $J_{c2}\simeq 0.241$.

Moreover, by using the entanglement excitation energy and a universal
frustration measure we considered the frustration properties of the system
and find the nonlinear dependence of the GMQD on the frustration. The
measure of the frustration can also be employed to detect the QPTs in this
system.

We emphasize that the two-site GMQD can detect the QPTs in this Heisenberg
system with NNN interaction. Although the two-site GMQD approach is only
effective in the finite-size systems, it has practical significance for the
usual theoretical and experimental studies. The problems of the GMQD (or QD)
in other systems with QPTs are interesting and need further consideration.

\section*{Acknowledgements} This work was supported by the National Nature Science
Foundation of China with Grant No.11005027; the Natural Science Foundation
of Zhejiang Province with Grant No.\thinspace Y6090058; the Program for
HNUEYT with Grant No. 2011-01-011; the NFRPC through Grant No. 2012CB921602
and the NSFC through Grants No. 11025527 and No. 10935010.

\end{document}